\documentclass[journal]{IEEEtran}
\pdfoutput = 1
\usepackage{graphicx,subfigure}
\usepackage{epstopdf}
\usepackage{mathtools}
\usepackage{amsthm}
\usepackage{amssymb}
\usepackage{ragged2e}
\usepackage[colorlinks,linkcolor=red,anchorcolor=red,citecolor=red]{hyperref}
\usepackage{caption2}
\usepackage{mathptm}

\usepackage{cases}
\usepackage[lined,boxed,ruled]{algorithm2e}
\usepackage{multicol}
\usepackage{multirow}
\usepackage{fancyhdr}

\newtheorem{thm}{Theorem}
\newtheorem{lem}{Lemma}
\newtheorem{cor}{Corollary}

\theoremstyle{definition}
\newtheorem{definition}{Definition}

\ifCLASSINFOpdf
\else
\fi
\begin{document}
\title{Analyzing and Disentangling Interleaved Interrupt-driven IoT Programs}
\author{Yuxia~Sun,
        ~\IEEEmembership{Member,~IEEE,}
         Song~Guo,
        ~\IEEEmembership{Senior~Member,~IEEE,}

        Shing-Chi~Cheung,
        ~\IEEEmembership{Senior~Member,~IEEE,}
        and Yong~Tang,
        ~\IEEEmembership{Member,~IEEE}
\thanks{This work was supported by the National Natural Science Foundation (Grant Nos. 61402197 and 61772211) of China, and Guangdong Province Science and Technology Plan Project (Grant No. 2017A040405030) in China.}
\thanks{Yuxia Sun is with the Department of Computer Science, Jinan University, Guangzhou 510632, China. Email: tyxsun@email.jnu.edu.cn.}
\thanks{S. Guo is with the Department of Computing, The Hong Kong Polytechnic, Hung Hom, Kowloon, Hong Kong SAR. Email: song.guo@polyu.edu.hk.}
\thanks{S. C. Cheung is with the Department of Computer Science, Hong Kong University of Science and Technology, Clear Water Bay, Hong Kong. Email: sccheung@cs.ust.hk.}
\thanks{Y. Tang  is with South China Normal University, Guangzhou 510631 Email: ytang@scnu.edu.cn.}}

\markboth{Submitted ,~Vol.~xx, No.~x, xx~20xx}%
{Shell \MakeLowercase{\textit{et al.}}:Analyzing and Disentangling Interleaved Interrupt-driven IoT Programs}

\maketitle

\begin{abstract}
In  the Internet of Things (IoT) community, Wireless Sensor Network (WSN) is a key technique  to enable ubiquitous sensing of environments and provide reliable services to  applications. WSN programs, typically interrupt-driven, implement the functionalities via the collaboration of Interrupt Procedure Instances (IPIs, namely executions of interrupt processing logic). However, due to the complicated concurrency model of WSN programs,  the IPIs are interleaved intricately and the program behaviours are hard to predicate from the source codes. Thus, to improve the software quality of WSN programs, it is significant to disentangle the interleaved executions and  develop various IPI-based program analysis techniques, including offline and online ones. As the common foundation of those techniques, a generic efficient and real-time algorithm to identify IPIs is urgently desired. However, the existing instance-identification approach cannot satisfy the desires. In this paper, we first formally define the concept of IPI. Next, we propose a generic IPI-identification algorithm, and prove its correctness, real-time and efficiency. We also conduct comparison experiments to illustrate that our algorithm is more efficient than the existing one in terms of both time and space. As the theoretical analyses and empirical studies exhibit, our algorithm provides the groundwork for IPI-based analyses of WSN programs in IoT environment.
\end{abstract}


\section{Introduction}\label{sec:introduction}
\IEEEPARstart{W}{ireless} Sensor Networks (WSNs), as essential components in the Internet of Things (IoT) ecosystem, have been increasingly employed for various applications. They help human beings enhance perception \cite{Homewood2013}, improve health \cite{Kim2011}, conserve environments \cite{Ramesh2014} and so on \cite{Shah2016}. To provide instant responses and save energy, programs running on WSN nodes are typically interrupt-driven with little energy consumption. The interrupt-driven concurrency mechanism of WSN programs involves both interrupt preemption and task scheduling, causing the interrupt-induced executions of the programs intricate. For example, in TinyOS programs (a group of mainstream WSN programs), an interrupt processing logic (called an interrupt-procedure) consists of one interrupt handler which is to be immediately performed and several interrupt-processing tasks whose executions are deferred. Because the executions of interrupt procedures, called the Interrupt Procedure Instances (IPIs), are interleaved in a complicated and unpredictable way, unexpected or even wrong instance interleaving is always inevitable during the executions of WSN programs. As a result, although the source programs seem short and simple, the program behaviors are difficult to predict, hard to test, and thus error-prone.

In recent years, researchers have reported various software faults in WSN programs \cite{Barrenetxea2008}, \cite{Raposo2017}, \cite{Schoofs2012}, \cite{Werner-Allen2006}. According to industrial remarks, the issues of software reliability have hampered the application of WSNs \cite{IoT}. Obviously, the software quality of WSNs has become critical concerns in the IoT community. However, analyzing and testing WSN programs is challenging, in that the development paradigm and tools are different from the traditional ones, and the interleaved program behaviors are too intricate to foresee from the codes using static analyses \cite{Larrucea2017}. In contrast to static analyses, dynamic analyses can precisely examine the actual program-behavior information obtained during program executions. Because WSN program behaviours consist of collaborative IPIs,  IPI-based program analyses are indispensable dynamic analysis techniques. IPI-based analyses of WSN programs can be classified into online analyses and offline analyses: both collect program-behavior information at run-time; the former analyzes at run-time, while the latter analyzes after the program terminates \cite{Dwyer2007}.
For conventional programs, online analyses has been shown  efficacious to uncover time-related issues such as concurrency bugs \cite{Park2015}, \cite{Roemer2018}, violations to temporal sequencing constraints \cite{Camilli2017}, performance issues \cite{Li2018}, and so forth. Due to the ability to timely generate analysis results, IPI-based online analyses also have the potentials to reveal time-related issues for WSN programs. In conclusion, to relieve the quality issues in WSN programs, it is significant to develop various IPI-based analysis techniques, including online and offline ones. Therefore, as the common foundation of all IPI-based analysis techniques (e.g. IPI-based profiling and testing) of WSN programs,  a generic IPI-identification algorithm is urgently desired.

In this paper, we aim to propose a generic algorithm to identify IPIs of TinyOS programs, which can support both online and offline analyses. Our research is enlightened by the pioneering work, namely Sentomist \cite{Zhou2010} and T-Morph \cite{T-Morph}, of testing TinyOS programs based on event-procedure-instances. Sentomist and T-Morph utilize an instance-identification algorithm with an implicit assumption that all task-posting operations of the tested program are atomic. In other words, event-procedure-instances are IPIs that involve no failed task-postings. However, most WSN applications allow a task-posting operation to be interrupted in the middle of its execution. Thus, it is necessary to relax the above atomicity assumption of the existing techniques and develop a more generic instance-identification algorithm.

Because Sentomist and T-Morph aim to support offline analysis, the issues of efficiency and real-time are not the major concerns of their instance-identification approach. In our study, we find it important to develop an efficient instance-identification algorithm to support dynamic analyses of WSN applications. This is because for dynamic online-analyses, efficient instance-identification is the foundation to enable efficient instance-based analyses. Even for dynamic offline-analyses, the collection of instance-based program behaviors also desires for efficient instance-identification to support efficient online collection. WSN applications are typically long-running programs, and thus always require long-running testing. The space and time overheads of conventional inefficient instance-identification will rise rapidly and ceaselessly with the running time, and thus disable long-running testing based on instances.

Moreover, we also find it necessary to propose a real-time instance-identification approach for dynamic analyses of WSN applications for the following observations: The instance-identification defective approach \cite{Zhou2010}, \cite{T-Morph} cannot determine all instance points at real-time, but having to postpone the determination of some instance-points, e.g., end-points, to the future. As a result, whenever a possible end-point of an instance is found, a program profiler must mark this point, keep associating the runtime-information to the instance which has possibly already been ended, and roll back to this point in the future when finding that this point is a real end-point. Such marking and rollbacks bring tight coupling between the information-collection logic of a dynamic analysis and the instance-identification logic, causing the dynamic analysis excessively complicated and error-prone. In addition, delayed instance-identification discourages instance-based online analyses from timely producing analysis results. Consequently, due to its defective instance-identification, a delayed profiling and testing approach for WSN programs cannot find concurrency bugs among instances in real-time or detect violations to real-time properties of instances.

 In this paper, to overcome the limitations of the existing instance-identification approach, we develop a novel instance-identification algorithm to facilitate IPI-based analyses of TinyOS programs. Firstly, our IPI-identification algorithm is generic without assuming atomic task-posting operations and without requiring IPI-based information-collection to know the algorithm¡¯s internal logic. Secondly, the IPI-identification algorithm has low overheads on time and space, and thus enable efficient IPI-based analyses. Thirdly, the algorithm identifies each IPI point of the program at real-time (i.e. immediately after the point occurs), which enables real-time collection of IPI-based program behaviors and makes online analyses possible. We prove the correctness, efficiency and real-time of our IPI-identification algorithm. Furthermore, we implement the prototype of our IPI-identification algorithm, and empirically compare its efficiency to the existing approach.

In summary, this paper makes the following contributions:

(1) \emph{Present a formal definition of Interrupt Procedure Instance (IPI) for WSN programs.} The implication of IPI is clarified and illustrated.

(2) \emph{Propose a generic algorithm for identifying IPIs of WSN programs.} The algorithm relaxes the assumption of atomic tasking-posting operations, and decouples its internal logic from the logics of various IPI-based analyses.

(3) \emph{Prove the correctness, efficiency and real-time of the IPI-identification algorithm theoretically.} The algorithm can be the common foundation of various IPI-based analyses of WSN programs, e.g., IPI-based profiling and testing.

(4)  \emph{Implement a prototype of our IPI-identification algorithm, and conduct comparison experiments to illustrate that our instance-identification approach excels the existing one on the running overheads of the analyzed program.}

We organize the rest of the paper as follows. Section \ref{sec:interruptprocedureinstances} outlines the fundamentals of TinyOS programming relevant to this paper and then presents the formal definition of IPIs. Our IPI-identification algorithm is depicted in Section \ref{sec:ipi-identificationalgorithm}. Section \ref{sec:algorithmanalysis} proves the correctness and real-time of our IPI-identification algorithm, and analyzes its time and space complexity. We experimentally compare the time and space costs of our algorithm to those of the existing instance-identification algorithm in Section \ref{sec:experimentstudy}. Section \ref{sec:relatedwork} reviews related work and Section \ref{sec:conclusionandfuturework} provides further discussions and summarizes this paper.

\section{Interrupt Procedure Instances}\label{sec:interruptprocedureinstances}

\subsection{Fundamentals of  WSN Programming}\label{subsec:fundamentals}
The concurrency model of such WSN operating systems as TinyOS is featured by the preemption execution of interrupt handlers and the delayed execution of tasks. TinyOS \cite{Levis2005}, written in nesC \cite{Gay2014}, is one of the mainstream operating system for WSN programming \cite{Sugihara2008}.

In TinyOS programs, a module could be in the form of a nesC \emph{event}, a nesC \emph{command}, a C \emph{function}, an \emph{interrupt handler}, or a \emph{task} \cite{Levis2009}. The nesC tools preprocess nesC code into C code, and then compile the C code into the target machine code \cite{TinyOS-2.x.}. In a nesC module \emph{m}, a task $t()$ and its task posting statement \emph{post(t)} are compiled into two C functions: $taskName\$runTask()$ and $taskName\$postTask()$, respectively, where $taskName$ denotes \emph{m\$t}. The function $taskName\$postTask()$ calls the OS scheduler's postTask function, namely $schedulerBasicP\$TaskBasic\$postTask()$, to push the task to the OS task queue. If the task is successfully pushed, it will be scheduled in a FIFO manner by the TinyOS scheduler by calling the function $taskName\$runTask()$.

\subsection{Definitions}\label{subsec:definitions}
Next, we formally define interrupt procedure instances. Let $IH$ be the interrupt handler of an interrupt $i$.

~\
\begin{definition}
The \emph{interrupt-procedure} of $IH$ consists of the static codes of three nesC modules -- $IH$, the \emph{callees of} $IH$(or $i$), and the \emph{tasks of} $IH$ -- where

(1) A \emph{callee of} $IH$  is a function that is called by $IH$, a \emph{callee of} $IH$, or a \emph{task of} $IH$.

(2) A \emph{task of} $IH$ is a task that is posted by $IH$, a \emph{callee of} $IH$, or a \emph{task of} $IH$.

\end{definition}

\begin{figure*}[!thb]
   \centering
   \begin{center}
   \scriptsize
     \includegraphics*[width=5.5in]{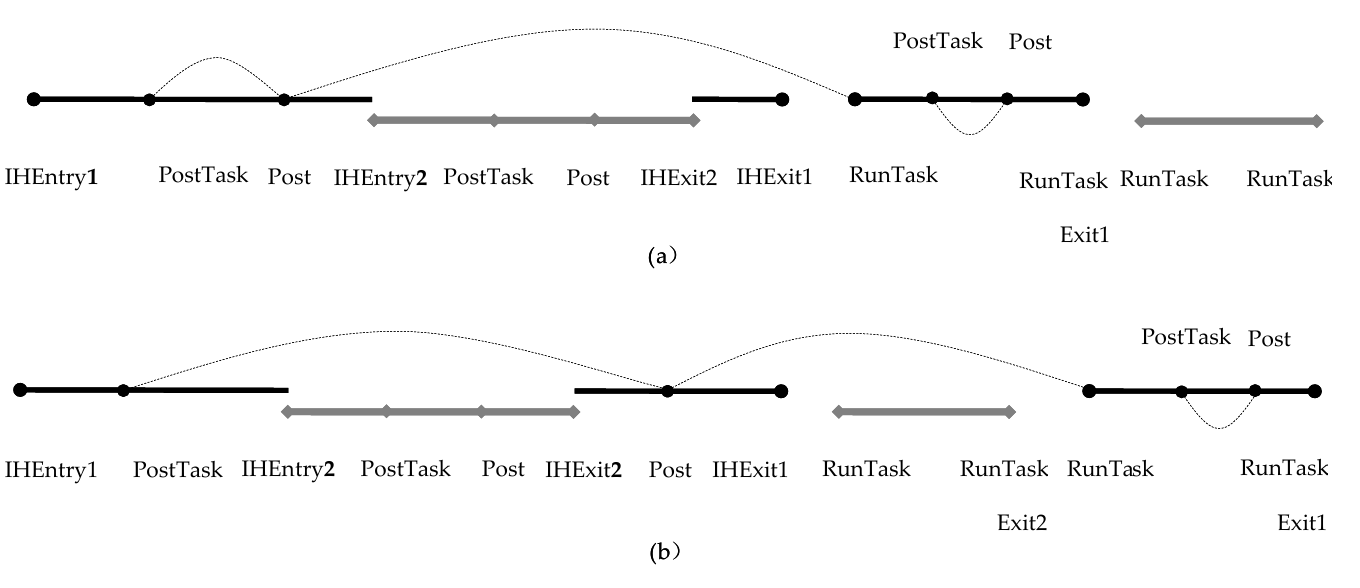}\\
   \caption{Examples of Interleaving IPIs}
   \label{fig:1}
   \end{center}
\end{figure*}

\begin{definition}
An \emph{interrupt-procedure-instance} (abbr. $IPI$) of $IH$(or $i$) is one execution of the interrupt procedure of $IH$. The \emph{callees} of the instance are the \emph{callees of} $IH$ that are executed in the instance. The \emph{tasks} of the instance are the tasks of $IH$ that are executed (i.e., successfully posted) in the instance.
\end{definition}
~\

Definition 1 is recursive. Note that when a task is successfully posted by a callee of an instance of $IH$, it also becomes a part of the instance. Therefore, we introduce ''the \emph{callees of } $IH$'' in Definition 1. Note also that a task posting is not necessarily successful because the OS task queue is a shared resource. For instance, in TinyOS2, a post will fail if the task is already in the task queue and has not started to execute \cite{Levis2009}. Thus, in Definition 2, we consider only the successfully posted tasks.

To illustrate the above definitions, we introduce notations for some execution points of IPIs in Table \ref{table:1}. Column 1 and 2 show the execution-points' types and their features, respectively. Intuitively, an IPI consists of an IH-running part and several (i.e., zero or more) task-running parts. The IH-running part, starting at the instance's \emph{IHEntry} point and ending at its \emph{IHExit} point, form an IHEntry-IHExit pair. A task-running part, starting at the task's \emph{RunTaskEntry} point and ending at its \emph{RunTaskExit} point, form a RunTaskEntry-RunTaskExit pair. Because tasks are scheduled in a FIFO manner, a RunTaskEntry-RunTaskExit pair will never contain other such pairs. Due to interrupt preemption, one instance's IHEntry-IHExit pair may embed into another instance's IHEntry-IHExit pair or RunTaskEntry-RunTaskExit pair.

\newcommand{\tabincell}[2]{\begin{tabular}{@{}#1@{}}#2\end{tabular}}
\begin{table}[!thb]
\centering
    \caption{Execution Point types of IPIs}
    \label{table:1}
        \begin{tabular}{ll}
          \hline\\[-2mm]
          \textbf{Execution-point type} & \textbf{Description} \\
          \hline
          \vspace{1mm}\\[-3mm]
          IHEntry & Entry of an interrupt handler \\
          \hline
          \vspace{1mm}\\[-3mm]
          IHExit  & Exit of an interrupt handler \\
          \hline
          \vspace{1mm}\\[-3mm]
          RunTaskEntry  &  \tabincell{l}{Entry of a \emph{taskName}\$runTask(), where \\ \emph{taskName} is a complete task name in \\ post-compiling format}\\
          \hline
          \vspace{1mm}\\[-3mm]
          RunTaskExit   &   \tabincell{l}{Exit of a \emph{taskName}\$runTask(), where \\ \emph{taskName} is a complete task name in \\ post-compiling format}\\
          \hline
          \vspace{1mm}\\[-3mm]
          PostTaskEntry &   \tabincell{l}{Entry of a \emph{taskName}\$postTask() }\\
          \hline
          \vspace{1mm}\\[-3mm]
          PostOk      &   \tabincell{l}{Point indicating a successful task posting \\ to the system task queue}\\
          \hline
          \vspace{1mm}\\[-3mm]
          PostFail    &	\tabincell{l}{Point indicating a failed task posting to \\ the system task queue}\\
          \hline
        \end{tabular}
    \end{table}

Figure \ref{fig:1}(a) illustrates an IPI, called IPI1, by black thick lines ended with solid circles. IPI1 starts at \emph{IHEntry1}, pauses because of the preemption execution of another instance named IPI2 (denoted by grey thick lines ended with solid diamonds), resumes after the preemption execution, pauses at \emph{IHExit1} due to the completed execution of IPI1's interrupt handler, resumes at \emph{RunTaskEntry1} due to the system task scheduling, and ends at \emph{RunTaskExit1} with no \emph{pending tasks} (i.e., the tasks that have been successfully posted to the OS task-queue but not yet scheduled to run). IPI1 posts two tasks: (1) the task posted at \emph{PostTaskEntry1} is successfully posted to the OS task-queue at \emph{PostOk1}, and scheduled to run at \emph{RunTaskEntry1}; (2) the task posted at \emph{PostTaskEntry1a} is unsuccessfully posted at \emph{PostFail1a}. The execution points of the same task are connected using continuous dotted lines in the figure, to show their \emph{corresponding} relationship. For example, \emph{PostTaskEntry1} and \emph{PostOk1} are the corresponding points of \emph{RunTaskEntry1}.
Figure \ref{fig:1}(b) shows another case of IPI1 when its task-posting procedure is interrupted by IPI2. Because IPI2'task is successfully posted prior to IPI1's task (namely \emph{PostOk2} occurs before \emph{PostOk1}), IPI2'task is also executed prior to IPI1's task (namely \emph{RunTaskEntry2} happens before \emph{RunTaskEntry1}).

Intuitively, an IPI starts at its \emph{IHEntry} point. If an instance has no successfully posted tasks, the instance ends at its \emph{IHExit} point; otherwise, it pauses at its \emph{IHExit} point (or at its \emph{RunTaskExit} points except the last one), resumes at its \emph{RunTaskEntry} points, and ends at its last \emph{RunTaskExit} point. If an instance is preempted by another one, the former instance pauses at the preempting instance's \emph{IHEntry} point, and resumes at the preempting instance's \emph{IHExit} point.

\section{IPI-identification Algorithm}\label{sec:ipi-identificationalgorithm}
In this section, we will propose a novel IPI-identification algorithm to overcome the limitations of the existing instance-identification technique described in Section \ref{sec:introduction}. We will firstly present the key execution points to identify instances, and then elaborate the instance-identification algorithm with theoretical analysis, and finally compare our algorithm to the existing one on time and space overheads with experiments.

\subsection{Key Execution Points}\label{subsec:keyexecutionpoints}
Our IPI-identification algorithm monitors the program execution at instruction-level and traces five types of key execution points, namely \emph{IHEntry}, \emph{IHExit}, \emph{RunTaskEntry}, \emph{RunTaskExit}, and \emph{PostOk} points. The first four points are used to trace the switches among instances (as explained in the rest of this subsection); the \emph{PostOk} points are utilized to identify each \emph{RunTaskEntry} point's instance and each instance's end-point (as detailed in Subsection \ref{subsec:algorithm})

During a run of a TinyOS program, such system operations as system initialization and system scheduling between task-executions are not driven by interrupts. Thus, the operations do not belong to any IPI, and can be regarded to belong to a specific \emph{Non-interrupt-instance}. Accordingly, when a program is launched and performs initialization, its execution belongs to the Non-interrupt-instance.

After the initialization, a program's execution might switch instances in the following four scenarios:

(1) At an IHEntry point: The currently executed instruction is the entrance instruction of an interrupt handler, which means that an interrupt just occurred and the program just started to execute the corresponding IPI (i.e., the IHEntry point's instance). Thus, at an IHEntry point, the program's execution switches to the IHEntry point's instance.

(2) At an IHExit point: The currently executed instruction is the exit instruction of an interrupt handler, and the next executed instruction will be the one preempted by the interrupt handler. Thus, at the immediate successor of an IHExit point, the program's execution switches to the instance preempted by the interrupt handler, and here the instance might be the Non-interrupt-instance or another IPI.

(3) At a RunTaskEntry point: The program starts to execute the task's function, namely \emph{taskName}\$runTaskEntry(). Thus, the program's execution switches to the instance of the RunTaskEntry point, namely the task's instance.

(4) At a RunTaskExit point: The currently executed instruction is the exit instruction of the task's function \emph{taskName}\$runTaskEntry(), and the next executed instruction will be an instruction in the system's task-scheduling function. Thus, at the immediate successor of a RunTaskExit point, the program's execution switches to the Non-interrupt-instance.

Next, we will prove that the above four scenarios contain all the possible cases for instance switches.

\begin{thm}{}
During the execution of a TinyOS program, instance switches only occur in one of the following execution points: IHEntry points, immediate successor points of IHExit points, RunTaskEntry points, and immediate successor points of RunTaskExit points.
\end{thm}

\begin{proof}
When a running TinyOS program switches instances, it switches into either an IPI or the Non-interrupt-instance, detailed as follows:

(1) The program-execution switches into an IPI only in one of the following three cases: (a) An interrupt occurs, and the program starts to execute the IPI's interrupt-handler; (b) A task-scheduling occurs, and the program starts to execute a task function of the IPI; (c) The execution of an interrupt-handler that previously preempted an IPI is ended, and the program continues to execute the IPI. In the above three cases, instance switches occur at the following three types of execution points, respectively: \emph{IHEntry} points, \emph{RunTaskEntry} points, and immediate successor points of \emph{IHExit} points.

(2) The program-execution switches into the Non-interrupt-instance only in one of the following two cases: (a) The execution of the interrupt-handler that previously preempted the Non-interrupt-instance is ended, and the program continues to execute the Non-interrupt-instance. (b) The execution of a task function of an IPI is ended, and the program continues to execute the Non-interrupt-instance. In the above two cases, instance switches occur at the following two types of execution points, respectively: immediate successor points of \emph{IHExit} points and immediate successor points of \emph{RunTaskExit} points.

Based on the above (1) and (2), Theorem 1 is proved.
\end{proof}

\subsection{Algorithm}\label{subsec:algorithm}

\begin{algorithm*}[!thb]
\label{alg:algrithm1}
\caption{InstanceIdentify (i)}
\LinesNumbered

/** Data structure: \\
INST: $\left \langle id, type \right \rangle$ --- an event-procedure instance, where non-zero integer id is the instance id; non-zero integer type is the instance type, which is interrupt number of the instance's triggering interrupt. \\
enum POSTYPE {START, END, INTERM} --- an instruction's position type in its instance, indicating that the instruction is a start point, an endpoint, or an intermediate point of its instance. \\
**/ \\
\textbf{Global}:  \quad \ $curInst$: INST  \qquad \qquad \qquad \qquad \qquad \qquad \qquad \, /* i's instance, initialized as $\langle 0,0 \rangle$ */  \\
\qquad \qquad \quad $instNum$: instance counter \qquad \qquad \qquad \qquad \qquad \qquad \quad \, \ \ /* initialized as 0 */  \\
\qquad \qquad \quad $pInst\_S$: preempted instances by IHs, stack of INST	\qquad \,  /* initialized as NULL */ \\
\qquad \qquad \quad $okInst\_Q$: pending tasks' instances, queue of INST	\qquad \ \ \ /* initialized as NULL */ \\
\textbf{Local}:  \qquad $instAfterExit$: INST  \qquad \qquad \qquad /* next instruction's instance that is different from i's instance*/  \\
\qquad \qquad \quad $curPos$: POSTYPE \qquad \qquad \qquad \qquad \qquad \qquad \, /* i's position type in its instance */  \\
\KwIn{\qquad $i$:  current instruction that is being executed}
\KwOut{\quad $curInst, curPos$}
\Begin{
$instAfterExit$ $\leftarrow$ NULL \qquad \qquad \qquad \qquad \qquad \qquad \quad /* NULL means instAfterExit is not set yet*/ \\
$curPos$ $\leftarrow$ INTERM \qquad \qquad \qquad \qquad \qquad \qquad \qquad \ \ /* i's default position type in its instance */ \\
$curInst$ $\leftarrow$ $\langle 0, 0 \rangle$; \qquad \qquad \qquad \qquad \qquad \qquad \qquad \qquad /* $i$ is of the non-interrupt-handling instance */ \\
    $instNum$ $\leftarrow$ 0; \\
    $pInst\_S$ $\leftarrow$ NULL;   $okInst\_Q$ $\leftarrow$ NULL; \\
\Switch{$i$'stype is:}{
    \Case{IHEntry:}{
        $pInst\_S$.push($curInst$);  \qquad \qquad \qquad \qquad \qquad \, /* save current instance to $pInst\_S$  */  \\
        increase $instNum$ by 1; \\
        $curInst$ $\leftarrow$ $\langle instNum$, IH's interrupt number $\rangle$; /* create a new instance */ \\
        $curPos$ $\leftarrow$ START; \qquad \qquad \qquad \qquad \qquad \qquad /* $i$ is the start point of its instance */ \\
    }
    \Case{IHExit:}{
        \If {($\neg$ $okInst\_Q$.contains($curInst$) )}{
            $curPos$ $\leftarrow$ END;  \qquad \qquad \qquad \qquad \qquad \quad \, /* i is the endpoint of its instance */
        }
        $instAfterExit$ $\leftarrow$ $pInst\_S$.pop();			\qquad \qquad \qquad \, /* next instance is the preempted instance retrieved */ \\
    }
    \Case(\qquad \qquad \qquad \qquad \qquad \qquad \qquad \quad \, /* $i$ is a successful task-posting point */){PostOk:}{
        $okInst\_Q$.add ($curInst$); \qquad \qquad \qquad \qquad \quad \, \ /* save PostOk's instance, also the task's instance */
    }
    \Case{RunTaskEntry:}{
        $curInst$ $\leftarrow$ $okInst\_Q$.remove();      		\quad \qquad \qquad \qquad /* get the task's instance */
    }
    \Case{RunTaskExit:}{
        \If {($\neg$ $okInst\_Q$.contains($curInst$) )}{
            $curPos$ $\leftarrow$ END;  \quad \qquad \qquad \qquad \qquad \qquad \ /* i is the endpoint of the current instance */
        }
        $instAfterExit$ $\leftarrow$ $\langle 0, 0 \rangle$; \quad \qquad \qquad \qquad \qquad \ \ \ /* next instruction is of the non-interrupt-handling instance */ \\
    }
}
\textbf{output} $curInst$, $curPos$;\qquad \qquad \qquad \qquad \qquad \qquad \quad \ /* i's instance, and i's position type in its instance */ \\
\If(\ \  \, /* instance-switch occurs, from i's instance */){($i$'s type==IHExit $\Vert$ $i$'s type==RunTaskExit)}{
     $curInst$ $\leftarrow$ $instAfterExit$;                                \qquad \qquad \qquad \qquad \quad \quad \ \ /* update current instance with next instance */
}}
\end{algorithm*}

Algorithm \ref{alg:algrithm1} shows our IPI-identification algorithm. It fires after each instruction $i$ is executed.  The algorithm inputs the instruction $i$, and outputs $i$'s instance (i.e.,\emph{curInst}) as well as $i$'s position in its instance (i.e., \emph{curPos}) at line 28. It reports three types of instruction positions, namely START, END and INTERM, indicating that the instruction is a start point, an endpoint or an intermediate point in its instance (line 2).

Algorithm \ref{alg:algrithm1} primarily utilizes the following data structure:

(1) The algorithm uses an INST $\langle id, type \rangle$ structure (line 1) to store an instance's information, where both \emph{id} and \emph{type} fields are non-zero. It uses a global \emph{instNum} to count and number all the instances (lines 8, 12 and 13). It also uses a special INST value $\langle 0, 0 \rangle$ to denote a Non-interrupt-instance. Thus, for an instruction that is not part of any instance, the algorithm sets its INST value with $\langle 0, 0 \rangle$ (lines 7 and 26). The algorithm uses the POSTYPE type (line 2) to define local \emph{curPos}. It sets the default value of \emph{curPos} as INTERIM (line 4); resets the value with STRAT when \emph{i} is an IHEntry point (line 14), or to END when \emph{i} is an instance endpoint (lines 16-17, 24-25).

(2) Because the execution of a tested program switches from an IHExit point (or a RunTaskExit point) into the instance of the point's immediate successor, our algorithm utilizes a local \emph{instAfterExit} to denote the instance's information and initializes \emph{instAfterExit} to NULL (line 3). When $i$ is an IHExit point (or a RunTaskExit point), the algorithm sets \emph{instAfterExit} with the instance information of the immediate successor of \emph{i} (lines 18, 26), and updates \emph{curInst} with \emph{instAfterExit} after outputting the instance information of $i$ (lines 28-30).

(3) Because the IH parts of multiple IPIs might be in multi-level nesting, our algorithm introduces a global INST stack, \emph{pInst\_S}, to trace the information of each instance preempted by interrupts. At each IHEntry point, it pushes the pre-updated curInst value into \emph{pInst\_S} (lines 10-11), which denotes the instance preempted by the IH. At each IHExit point, it pops the INST value from \emph{pInst\_S}, and updates \emph{instAfterExit} with the value (lines 15, 18). This value represents the instance of the immediate successor of the IHExit point (as Lemma 1 exhibits in Section \ref{sec:algorithmanalysis}).

(4) Because TinyOS uses a system task-queue to schedule the successfully posted tasks, our algorithm also introduces a global INST queue, \emph{okInst\_Q}, to trace the instance of each successfully posted task. At each PostOk point, it adds the curInst value to \emph{okInst\_Q} (lines 19-20), and the value represents the PostOk point's instance, namely the instance of the pending task successfully posted at the PostOk point. At each RunTaskEntry point, the algorithm removes the first value from \emph{okInst\_Q} (lines 21-22). The removed value denotes the running task's instance, namely the RunTaskEntry point's instance (as Lemma 2 reveals in Section \ref{sec:algorithmanalysis}).

Next, we depict how Algorithm \ref{alg:algrithm1} traces the instance switches by setting the global \emph{curInst}. When the tested program starts to run, the algorithm initializes \emph{curInst} with $\langle 0, 0 \rangle$ (line 7), denoting current instance is No-interrupt-instance. The algorithm updates the value of \emph{curInst} at the following key execution-points:

(1) When \emph{i} is an IHEntry point, the algorithm creates an INST value using the interrupt number of IH and the current instance number to denote \emph{i}'s instance, and updates \emph{curInst} with the value (line 13).

(2) When \emph{i} is an IHExit point, the algorithm pops an INST value from \emph{pInst\_S}, sets \emph{instAfterExit} to the value (line18), and updates \emph{curInst} with the value after outputting \emph{i}'s instance \emph{curInst} (lines 28-30).

(3) When \emph{i} is a RunTaskExit point, the algorithm sets \emph{instAfterExit} to Non-interrupt-instance (line 26), and updates \emph{curInst} with the value after outputting \emph{curInst} (lines 28-30).

(4) When \emph{i} is a RunTaskEntry point, the algorithm removes the first value from \emph{okInst\_Q}, and updates \emph{curInst} with the value (line 22).

Finally, we address how Algorithm \ref{alg:algrithm1} finds out an instance-endpoint by setting the local \emph{curPos}. The algorithm initializes \emph{curPos} with default INTERM (line 4), indicating the instruction \emph{i} is neither a start-point nor an end-point of its instance. When \emph{i} is an IHEntry point, it sets \emph{curPos} with START (line 14). When \emph{i} is an IHExit or RunTaskExit point, the algorithm checks whether or not the INST value of the point's instance is in \emph{okInst\_Q}; if not, sets \emph{curPos} to END (lines 15-17, 23-25), and the point is the end-point of its instance (as Lemma 3 shows in Section \ref{sec:algorithmanalysis}).

\section{Algorithm Analysis}\label{sec:algorithmanalysis}
In this section, we will theoretically analyze the correctness, real-time and efficiency of our IPI-identification algorithm.

\begin{lem}
Lemma 1. When Algorithm \ref{alg:algrithm1} is processing an IHExit execution point, the popped INST value from the stack \emph{pInst\_S} is the instance information of the immediate successor of the IHExit point.
\end{lem}

\begin{proof}
(1) At and only at each IHEntry point of the tested program, TinyOS pushes the interrupted site of the instruction preempted by the IH into the system stack, and at the same time Algorithm \ref{alg:algrithm1} pushes the instance information of the instruction to the algorithm stack \emph{pInst\_S}; (2) At and only at each IHExit point of the tested program, TinyOS pops the system stack, and at the same time Algorithm \ref{alg:algrithm1} pops the algorithm stack \emph{pInst\_S}. Obviously, the above two stacks synchronize on all the stack push and pop operations, and the top elements of the two stacks denote a same instruction all the time. For this reason, when Algorithm \ref{alg:algrithm1} is processing an IHExit execution point, the popped INST value from the stack \emph{pInst\_S} is the instance information of the instruction preempted by the IH, namely the instance information of the immediate successor of the IHExit point.
\end{proof}

\begin{lem}
When Algorithm \ref{alg:algrithm1} is processing a RunTaskEntry execution point, the removed INST value from the queue \emph{okInst\_Q} is the instance information of the immediate successor of the RunTaskEntry point.
\end{lem}

\begin{proof}
(1) At and only at each PostOk point of the tested program, TinyOS adds the entry address of the successfully posted task at the point to the system task queue, and simultaneously, Algorithm \ref{alg:algrithm1} adds the instance information of the task to the algorithm queue \emph{okInst\_Q}; (2) At and only at each RunTaskEntry point of the tested program, TinyOS dequeues the system queue and the removed element is the entry address of the currently running task, and simultaneously, Algorithm \ref{alg:algrithm1} dequeues the algorithm queue \emph{okInst\_Q}. Evidently, the above two queues act in the same pace on all the enqueueing and dequeueing operations, and the head elements of two queues represent a same task all the time. Thus, when Algorihtm \ref{alg:algrithm1} is processing a RunTaskEntry execution point, the dequeued INST value from the queue \emph{okInst\_Q} is the instance information of the currently running task, namely the instance information of the RunTaskEntry point.
\end{proof}

\begin{lem}
When a tested TinyOS program is executing an IHExit or RunTaskExit point, if the queue \emph{okInst\_Q} of Algorihtm \ref{alg:algrithm1} does not contain the point's instance information, the point is the endpoint of the instance.
\end{lem}

\begin{proof}
During the tested program is running, both Algorithm \ref{alg:algrithm1}'s queue \emph{okInst\_Q} and the TinyOS task queue are initialized to null, and then act in the same pace on all the enqueueing and dequeueing operation (as proved in Lemma 2). $\rightarrow $ There is a one-to-one mapping between the instance information of the tasks in \emph{okInst\_Q} and the entry addresses of the tasks in TinyOS task queue.$\rightarrow $ If at some moment, a given instance has no instance information in \emph{okInst\_Q}, then the instance has no pending tasks at that moment.$\rightarrow $ At an IHExit or RunTaskExit execution point of the tested program, if the instance of the point has no instance information in \emph{okInst\_Q}, then the instance has no pending tasks, and hence the IHExit or RunTaskExit point is the instance's endpoint.
\end{proof}

\begin{cor}
The IPI-identification of Algorithm \ref{alg:algrithm1} is correct and real-time.
\end{cor}

\begin{proof}
(1) By taking Theorem 1, Lemma 1 and Lemma2 together, the following conclusion can be drawn: Algorithm \ref{alg:algrithm1} traces all the instance switches and gets the instance information on each switch correctly; According to Lemma3, Algorithm \ref{alg:algrithm1} identifies the start point and endpoint of each instance correctly. Therefore, Algorithm \ref{alg:algrithm1} is correct. (2) For each executed instruction $i$, immediately before the next instruction is executed, Algorithm \ref{alg:algrithm1} can output $i$'s instance information and the type of $i$'s position in its instance. Therefore, Algorithm \ref{alg:algrithm1} is real-time.
\end{proof}

\begin{cor}
Both the space complexity and the time complexity of Algorithm \ref{alg:algrithm1} are constant $\mathcal{O}$(1).
\end{cor}

\begin{proof}
(1) Algorithm \ref{alg:algrithm1} utilizes a counter $instNum$, three variables (namely $curInst$, $newNextInst$, and $curPos$), a stack $pInst\_S$ and a queue $okInst\_Q$. The maximum stack depth is the maximum interrupt-nesting depth, which is a small constant in practice. The maximum size of the queue is the maximum size of the OS task queue. For example, in TinyOS1, the maximum size of the OS task-queue is 8, and in TinyOS2, although no size limitation (so as to avoid queue overflow), the maximum size is still a small constant in practice. Therefore, the space overhead of Algorithm \ref{alg:algrithm1} is $\Theta$(1).

(2) For each executed instruction, Algorithm \ref{alg:algrithm1} gets its execution-point type with a constant time, and processes the following five types of points: $IHEntry$, $IHExit$, $RunTaskEntry$, $RunTaskExit$ and $PostOk$. At each point above, our algorithm performs the following actions: one counter increment (line 12), one stack-push (stack-pop, enqueueing, dequeueing) operation (line 11, 18, 20 or 22), several assignments (lines 3-4, 7-9, 13-14, 17-18, 22, 25-26, and 30), several logic operations (lines 16, 24, and 29 ), and one queue searching operation (line 16 or 24). Because the maximum queue size is a small constant as described above, the maximum time overhead for processing each point is a constant. Therefore, the time complexity of our instance identification process is $\mathcal{O}$($n$), where $n$ is the total number of the executed instructions, and $n$ increases with the running time. Suppose that $t$ is the running time of the tested program, and that the program can execute up to N instructions per unit time, then $\mathcal{O}$($n$) = $\mathcal{O}$($t$*N) = $\mathcal{O}$($t$). Because the time for running a program once is limited, namely $t$$<$C (C is a large constant), $\mathcal{O}$($n$) = $\mathcal{O}$($t$) = $\mathcal{O}$(C) = $\mathcal{O}$(1). Therefore, the time overhead of Algorithm 1 is $\mathcal{O}$(1).
\end{proof}

\section{Experimental Study}\label{sec:experimentstudy}

In this section, we will empirically study the follow question on efficiency:

RQ: In practice, does our instance-identification approach excels the existing approach on the running overheads of the analyzed program?

\subsection{Experimental Setup}\label{subsce:experimentalsetup}
We implemented our instance-identification tool in Java by utilizing the probe mechanism of Avrora \cite{Titzer2005}, a cycle-accurate instruction-level simulator for sensor network. The probe fires when each instruction of the program-under-analysis is executed by the Avrora interpreter. The tool for implementing the existing instance-identification technique (called the old tool) is obtained by merely keeping the code of Sentomist (or T-Morph) tool \cite{Source} for instance identification.

We performed all the experiments on top of Avrora 1.7.113 \cite{CVS} with a simulated Mica2 platform and AT-Mega128 microcontroller. The underlying operation system is TinyOS 2.1. We installed TinyOS on the platform of Cygwin \cite{HoCp} and Windows XP. We ran our experiments on a desktop computer with a 2.7 GHz Intel dual-core processor and 1GB RAM.

\begin{table}[!thb]
\centering
    \caption{Subject programs and running settings}
    \label{table:2}
    \begin{tabular}{llll}
    \hline
    \textbf{Subject}  &  \textbf{RunGroup}  &  \textbf{Sampling period}  & \textbf{Node Monitored} \\
             &  \textbf{No.}       &  \textbf{(ms)}   &  \\
    \hline
    \multirow{2}{*}{Sub1} & R1  &100  & Source node  \\
    \cline{2-4}
                       & R2  &20  & Source node   \\
    \hline
    \multirow{2}{*}{Sub2} & R3  &100  & Source node  \\
    \cline{2-4}
                       & R4  &20  & Source node   \\
    \hline
    Sub3  &R5  & Default of Avrora &Source node   \\
    \hline
    \multirow{2}{*}{Sub4} & R6  &100  & Intermediate node  \\
    \cline{2-4}
                       & R7  &20  & Intermediate node   \\
    \hline
    \multirow{2}{*}{Sub5} & R8  &Set by TestCTP  &Benign node \\
    \cline{2-4}
                       & R9  &Set by TestCTP  & Buggy node\\
    \hline
\end{tabular}
\end{table}

Table \ref{table:2} lists the subject programs and their run settings in the experiments. The subjects are five variants of three typical WSN applications, namely Osilloscope, TestBlink and TestCTP \cite{Source}, \cite{TinyOS}. They cover three typical interrupts, namely ADC (Analog to Digital Conversion), SPI (Serial Peripheral Interface), and TIMER interrupts, respectively. Osilloscope is a sensor data collection program using single-hop packet transmissions. TestBlink implements multihop packet transmissions. TestCTP transports sensor readings using a routing protocol called Collection Tree Protocol (CTP) \cite{Gnawali2009}. Column 1 denotes the subject's name. Sub1-3 are three variants of Osilloscope. Sub4 is a variety of TestBlink. Sub5 is the TestCTP application in the Sentomist release package.

\begin{figure}[!ht]
  \centering
  \subfigure[Space overhead]{
    \label{fig:subfig:a} 
    \includegraphics[width=3.5in]{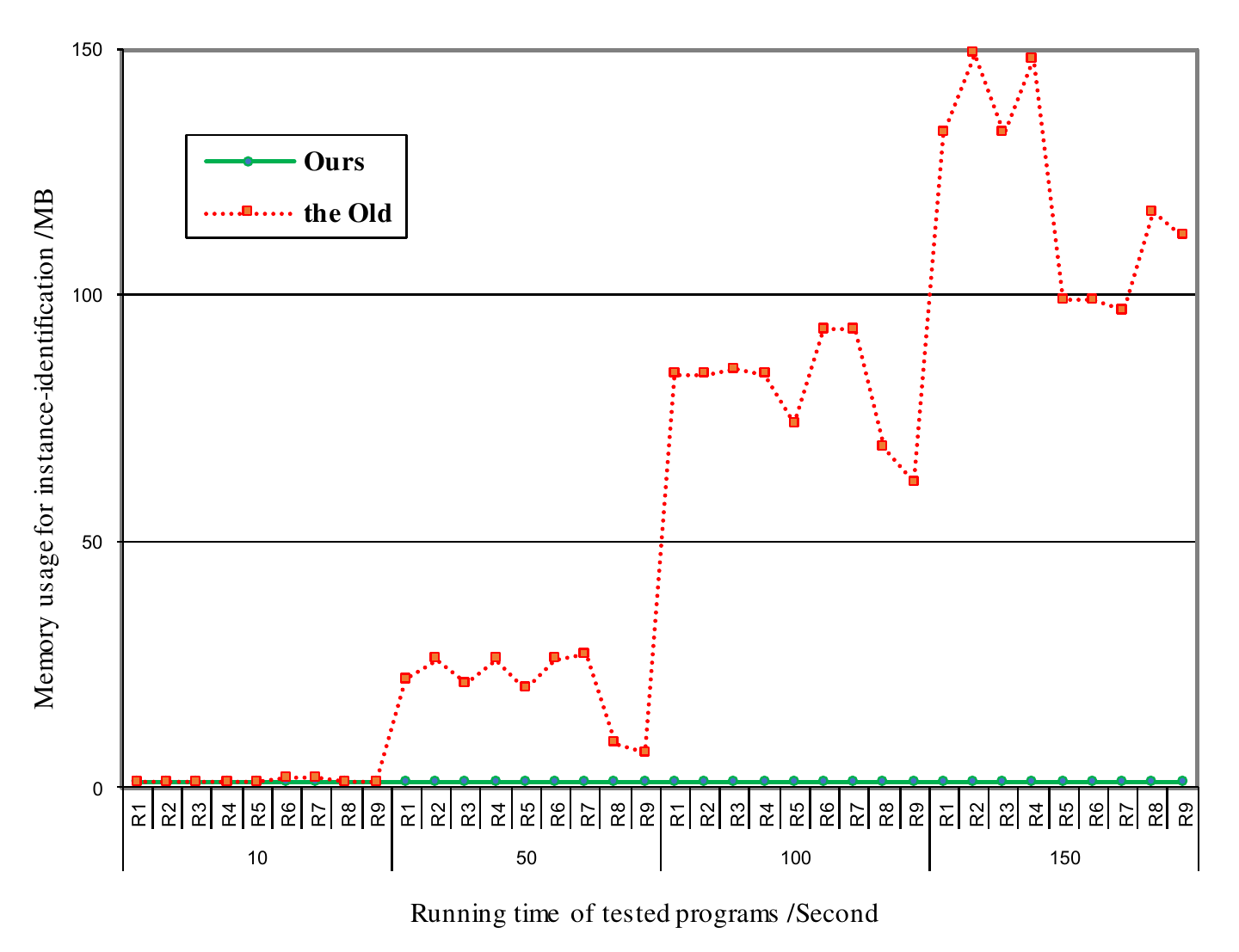}}
  \hspace{1in}
  \subfigure[Time overhead]{
    \label{fig:subfig:b} 
    \includegraphics[width=3.5in]{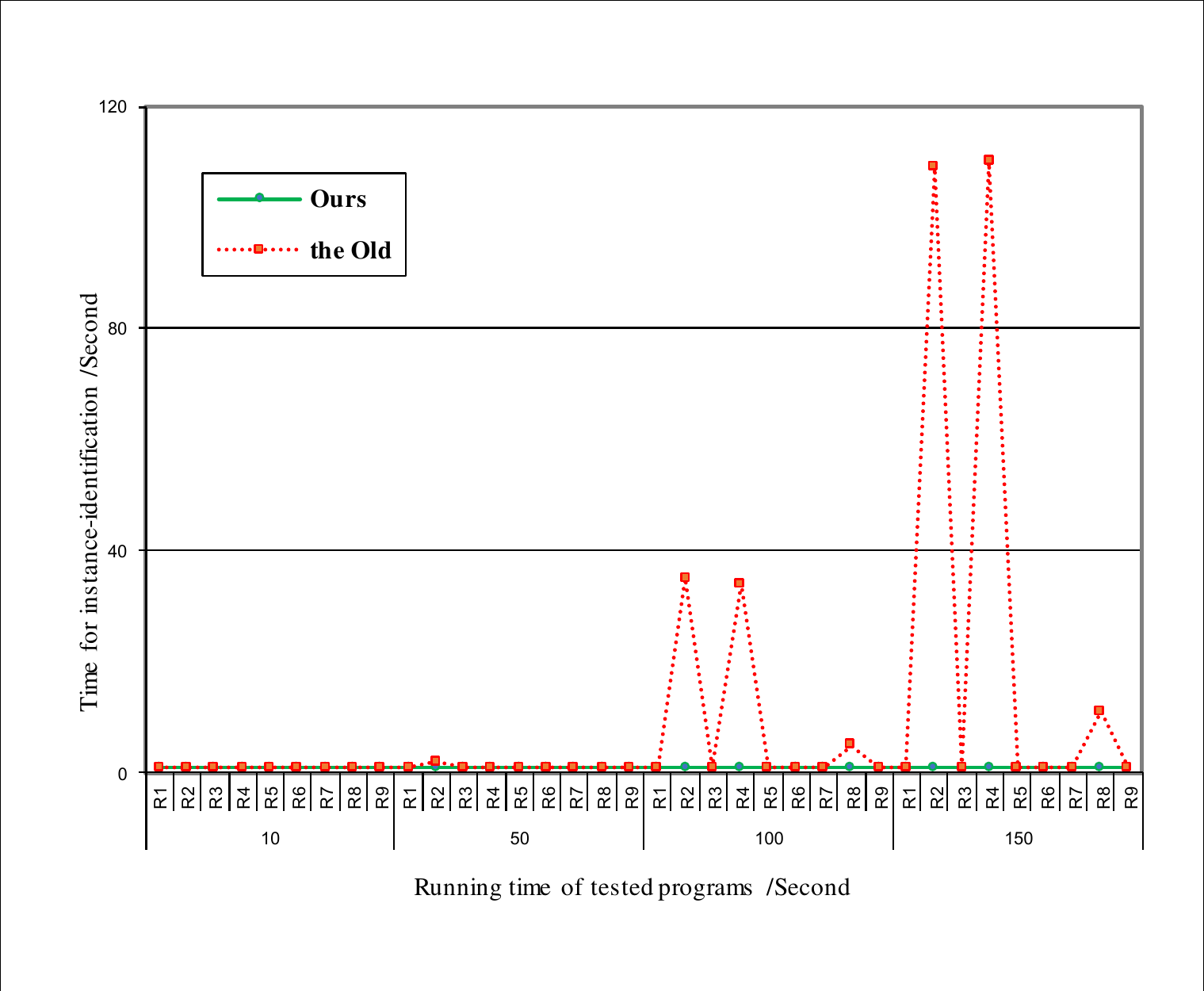}}
  \caption{Overheads of two instance-identification tools}
  \label{fig:2} 
\end{figure}

%

We monitored the running overheads of each subject on a single node, where different executions of a subject involve distinct running time and/or distinct source node's sampling periods. The overheads might go up with increasing running time. Column 2 reports nine test-run groups for the subjects, and each group $R_i$ (1 $\leq i \leq $ 9) consists of four test runs with four running time (measured in $seconds$): 10, 50, 100, and 150, respectively. Thus, Column 2 contains 9*4=36 test runs in all. The overheads on a node might also be affected by the source node's sampling period. Column 3 reports the source node's sampling period (measured in $millisecond$) in each test run group. For each subject whose sampling period is alterable (i.e. Sub1-2 and Sub4), we ran the subject twice with two sampling periods: one is longer and the other is shorter, respectively. The sampling period of Sub3 is determined by Avrora, and that of TestCTP is set by the implementation of TestCTP. Therefore, for Sub3 and Sub5, we utilized the default sampling period. Column 4 reports the monitored node in each test run group. In Sub5, there is a bug of stopping packet-sending. When the bug occurs, the number of concerned instances on the buggy node might stop increasing, and this might influence the overhead's increment with the running time. To observe the overheads of Sub5 with and without that possible influence, respectively, we monitored a benign node in R8 as well as the buggy node in R9.

\subsection{Experimental Results}\label{subsce:experimentalresults}
We applied our tool and the old tool of instance-identification, respectively, to all the test runs listed in Table \ref{table:2}, and measured each tool's run-time overheads for instance-identification. The results are shown in Figure \ref{fig:2}, where solid green lines represent our tool's overheads and dot red lines denote the old tool's overheads. In Figure \ref{fig:subfig:a}, each line dot denotes space cost for instance identification, and the dot's height expresses the memory usage in MB. In Figure \ref{fig:subfig:b}, each line dot indicates time overhead for instance identification, and the height expresses the execution time in seconds. Each test run occupies a position along the horizontal axis, and 36 test runs are classified into four groups based on four running time (i.e., 10, 50, 100, and 150 seconds). We make the following observation from the graphs in Figure \ref{fig:2}: using our tool, both space cost and time overhead for all test runs are small constants; in contrast, using the old tool, space cost for all test runs and time overhead for some test runs go up with increasing running time.

\section{Related Work}\label{sec:relatedwork}

\subsection{Instance-based Testing of WSN programs}\label{subsec:instance-basedtestingofWSNprograms}
By analyzing diverse run-time information of interrupt procedure instances in different ways, various instance-based profiling and testing techniques can be developed for WSN programs. Sentomist \cite{Zhou2010} and T-Morph \cite{T-Morph} are the pioneering instance-based testing approaches. Sentomist aims to find transient bugs in TinyOS programs. It online collects the instruction-coverage information during instance-intervals with vectors, and offline detects the outlier instance by vector mining. T-Morph detects bugs but not limited to transient ones. It online collects function-invocation sequences of instances, and offline analyzes the suspicious patterns among the sequences by tree mining.

Instances are triggered by interrupts. To generate random interrupts, Regehr [REGEHR 2005] proposes a random testing strategy. By utilizing this strategy, instance-based testing can permute the interleavings of instances. Like other dynamic testing of WSN programs, instance-based testing on real hardware is always difficult. This is because instrumentation may impact programs' behaviors, and hardware's internal states are always unaccessible by developers \cite{Kamph2010}, \cite{Yu2017}. WSN simulation allows more detailed inspection of program execution before deployment. Instruction-level simulators, such as Avrora \cite{Titzer2005} (AVR platform) and COOJA/MSPSim \cite{Eriksson2009} (MSP430 platform), can simulate motes running on different operating systems.  Other popular code-level simulators include TOSSIM \cite{Levis2003}, ATEMU \cite{Polley2004}, and so on. Although simulators can only simulate limited amount of hardware behaviors, they can be the most flexible way to analyze WSN programs dynamically. For example, Avrora contains a flexible framework for running and analyzing programs without changing the programs themselves, therefore instance-based testing tools can be conveniently constructed based on Avrora, as the tools of Sentomist and T-Morph show.

\subsection{Dynamic Analysis and Verification of IoT Programs}\label{subsec:dynamicanalysisandverificationofIoTprograms}
As IoT becomes increasingly pervasive, we need more and broader software engineering support to improve the quality of WSN-based IoT programs \cite{Eugster2015}. In recent years, apart from instance-based analysis, other dynamic analysis techniques have been developed for WSN applications: For example, Sundaram et al. propose an efficient approach to intra-procedural and inter-procedural control-flow tracing \cite{Sundaram2013}; Dylog \cite{Dong2016} provides a dynamic event-logging facility for networked embedded programs to support efficient and accurate analysis. Based on various runtime data logs, some testing techniques have been proposed for WSN programs: For instance, D2 \cite{Dong2013} employs function count profiling and PCA (Principal Component Analysis) to reveal network anomalies; Khan et al. applies discriminative sequence mining to uncover interactive bugs \cite{Khan2014}. There has been some work in runtime checking of WSN applications: For instance, nesCheck \cite{Midi2017} check errors violating memory safety and KleeNet \cite{Sasnauskas2010} uses symbolic analysis to find bugs.

\section{Conclusion and Future Work}\label{sec:conclusionandfuturework}
To relieve the quality issues of interrupt-driven WSN programs, it is essential to develop various profiling and testing techniques based on the program behaviours of IPIs. In this paper, we proffer the formal definition of IPI and expound its meanings. To support IPI-based analyses of TinyOS programs, we construct an IPI-identification algorithm, and  theoretically prove its correctness, efficiency and real-time. We also conduct comparison experiments to illustrate that the our instance-identification approach has lower running overheads than the existing one. In conclusion, we contribute a generic, efficient and realtime IPI-identification algorithm, building the firm base for IPI-based analyses of WSN program in IoT environment.

 Based on our IPI-identification algorithm, multifarious IPI-based profiling and testing techniques can be proposed for WSN programs. In the near future, we will study IPI-based bug patterns, and develop an IPI-based testing technique with the patterns for WSN-based IoT programs.

\section*{Acknowledgments}
The authors thank the anonymous reviewers for their insightful comments.

\bibliographystyle{IEEEtran}
\bibliography{RefBase}

\begin{IEEEbiography}[{\includegraphics[width=1in,height=1.25in,clip,keepaspectratio]{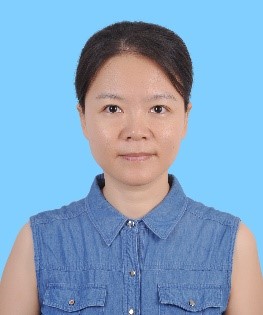}}]{Yuxia Sun}
received her B.S. degree from Huazhong University of Science and Technology, and the Ph.D. degree from Sun Yat-sen University, both from Department of Computer Science. She is currently an Associate Professor in the Department of Computer Science at Jinan University. She was a Research Associate at the Hong Kong Polytechnic University and at the University of Hong Kong, and a Research Scholar in the College of Computing at Georgia Institute of Technology. Her research focuses on software engineering, software safety and system safety.
\end{IEEEbiography}
\begin{IEEEbiography}[{\includegraphics[width=1in,height=1.25in,clip]{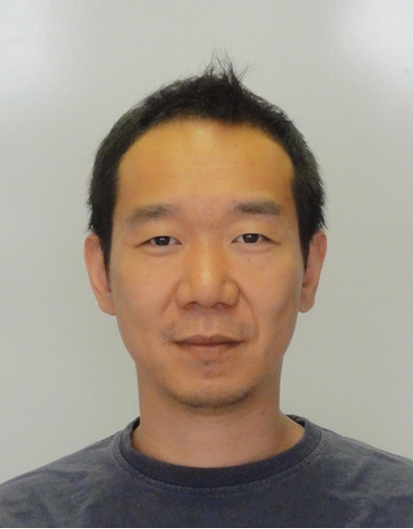}}]{Song Guo }
is a Full Professor at Department of Computing, The Hong Kong Polytechnic University. He received his Ph.D. in computer science from University of Ottawa and was a professor with the University of Aizu from 2007 to 2016. His research interests are mainly in the areas of big data, cloud computing and networking, and distributed systems with over 400 papers published in major conferences and journals. His work was recognized by the 2016 Annual Best of Computing: Notable Books and Articles in Computing in ACM Computing Reviews. He is the recipient of the 2017 IEEE Systems Journal Annual Best Paper Award and other five Best Paper Awards from IEEE/ACM conferences. Prof. Guo was an Associate Editor of IEEE Transactions on Parallel and Distributed Systems 2011-2015 and an IEEE ComSoc Distinguished Lecturer 2016-2017. He is now on the editorial boards of IEEE Transactions on Emerging Topics in Computing, IEEE Transactions on Sustainable Computing, IEEE Transactions on Green Communications and Networking, and IEEE Communications. Prof. Guo also served as General, TPC and Symposium Chair for numerous IEEE conferences. He currently is the Director of ComSoc Membership Services and Member of ComSoc Board of Governors. Prof. Guo has also served as General, TPC and Symposium Chair for numerous IEEE conferences.
\end{IEEEbiography}
\begin{IEEEbiography}[{\includegraphics[width=1in,height=1.25in,clip]{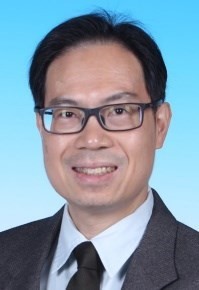}}]{Shing-Chi Cheung}
received his doctoral degree in Computing from the Imperial College London. In 1994, he joined The Hong Kong University of Science and Technology, where he is a full professor of Computer Science and Engineering. He participates actively in program and organizing committees of major international software engineering conferences. He was the General Chair of the 22nd ACM SIGSOFT International Symposium on the Foundations of Software Engineering (FSE 2014). He was a director of the Hong Kong R \& D Center for Logistics \& Supply Chain Management Enabling Technologies. His research interests include program analysis, testing and debugging, big data software, cloud computing, internet of things, and mining software repository.
\end{IEEEbiography}
\begin{IEEEbiography}[{\includegraphics[width=1in,height=1.25in,clip]{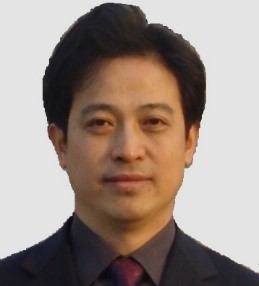}}]{Yong Tang}
got his BS and MSc degrees from Wuhan University in 1985 and 1990 respectively, and PhD degree from University of Science and Technology of China in 2001, all in computer science. He is now a Professor and Dean of the School of Computer Science at South China Normal University(SCNU). He serves as the Director of Services Computing Engineering Research Center of Guangdong Province. He was vice Dean of School of Information of Science and Technology at Sun Yat-Sen University, before he joined SCNU in 2009. He has published more than 200 papers and books. As a supervisor he has had more than 40 PhD students and Post Doc researchers since 2003 and more than 100 Master students since 1996. He is a Distinguished Member and the vice director of Technical Committee on Collaborative Computing of China Computer Federation (CCF). He has also served as general or program committee cochair of more than 10 conferences.
\end{IEEEbiography}

\end{document}